# Experimental Observation of Reversed Doppler Effects in Acoustic Metamaterials

**Shilong Zhai, Xiaopeng Zhao[1]\*, Song Liu, Chunrong Luo**

[1]Smart Materials Laboratory, Department of Applied Physics, Northwestern Polytechnical University, Xi'an 710129 P. R. China,
\*email: xpzhao@nwpu.edu.cn

This paper reports an experimental observation of broadband reversed Doppler effects using an acoustic metamaterial with seven "flute-like" double-meta-molecule clusters. Simulations and experiments verify that this locally resonant acoustic metamaterial with simultaneous negative elastic modulus and mass density can realize negative refraction in a broad frequency range. The constructed metamaterial exhibits broadband reversed Doppler effects. The frequency shift increases continuously as the frequency increases. The assembling of double-meta-molecule clusters introduces a new direction in designing double-negative acoustic metamaterials in an arbitrarily broad frequency range and other various applications.

The Doppler effect refers to the change in frequency from the wave source caused by the relative motion between the wave source and the observer[1, 2]. This phenomenon is applied in many fields, including scientific research, space technology, traffic management, and medical diagnosis[3-6]. In 1968, Veselago[7] theoretically predicted that metamaterials[8-10] with negative refractions[11, 12] can realize the reversed Doppler effect. Recent studies show that the reversed Doppler shift in electromagnetic waves can be achieved using transmission line[13-15], backward dipolar spin waves in a magnetic thin film[16, 17] and photonic crystals[18, 19]; in acoustics, phononic crystal[20] is a feasible material to obtain the reversed Doppler effect. However, this abnormal phenomenon appears experimentally only in the frequency range corresponding to the energy gap, and the band is narrow. Based on nonresonant elements, a quasi-1D double-negative acoustic metamaterial can also realize the reversed Doppler effect in the low-frequency region[21].

In many applications, such as metamaterial absorber and cloak, broadband metamaterials are required. Given that the mechanism of metamaterial is based on resonance in the periodic structure[22, 23], the bandwidth of this resonance is narrow by nature. To achieve broadband absorption, some projects employ nested elements, such as concentric square rings, and realize double- and even triple-band absorption[24, 25]; however, these frequency regions are not close enough to combine into a broader band because of the limitation of element size. Simulated studies have verified multiplexed configuration as a convenient approach to obtaining multiresonance and broadband absorption[26]. Xie et al. demonstrated that the resonating nature resulting from space folding instead of local resonance within the unit cell creates an extremely broad frequency field (i.e., more than 1000 Hz) of negative index[27]. Nonresonant metamaterial elements can also be used for broadband cloaks[28, 29]. Gradient-index structures can be utilized for broadband antireflection[30, 31] employing phononic crystals. However, to some extent, the multiple-scattering-based mechanism limits the working frequency region of acoustic waves in phononic crystals[32]. A general approach to realizing broadband effects is still required. However, a general method of designing and fabricating a controllable broadband double negative acoustic metamaterial is yet to be reported. Therefore, realizing arbitrarily broadband acoustic transmission and obtaining the reversed Doppler effect using negative refraction materials are difficult.

Electromagnetic waves consist of many photonics with different energies and frequencies. Visible white light (400 THz-800 THz) can be obtained via mixing disparate photons, with colors corresponding to red, orange, yellow, green, cyan, blue, and purple. Analogous to photons, surface plasmon polaritons exist in acoustic metamaterials, which can be generated by the resonance of a sort of acoustic meta-molecule[33]. The frequency of plasmon polaritons is related to the geometry of structural units and responds to the region where the simultaneous negative mass density, modulus, and the reversed Doppler effect of the meta-molecule material appear. Nevertheless, the frequency bandwidth is relatively narrow[33, 34]. Inspired by the fact that visible light is composed of seven-color lights, we propose a "flute-like" model of acoustic meta-molecule cluster containing seven double meta-molecules with different dimensions. A metamaterial sample is experimentally constructed based on this model. Transmission and reflection results are obtained via experimental measurements and numerical simulations, from which mass density and bulk modulus are derived to be simultaneously negative in a broadband range. The calculated and measured refractions are also negative in the broad resonant frequency region. Taking advantage of this broadband double-negative sample, we experimentally realize reversed Doppler effects from 1.186 KHz to 6.534 KHz.

The analogy betwwen acoustic surface plasmon polaritons and photons is shown in Fig. 1. The frequency $v$ of photon is determined by its energy $\varepsilon$, which is expressed as

$$\varepsilon = hv. \qquad (1)$$

No interactions occur among between photons. In the visible portion of the electromagnetic waves, photons with different frequencies correspond to lights with seven different colors. Visible white light is formed by mixing these seven colors. Acoustic metamaterials can generate oscillations of surface plasmon polaritons near the resonant frequency (from $\omega_0 - \delta\omega$ to $\omega_0 + \delta\omega$); such oscillations directly lead to the abnormal effective parameters of materials, with mass density and bulk modulus shown as follows[33]:

$$\rho_{eff1} = \rho_0 \left(1 - \frac{F_t \omega_1^2}{\omega^2 - \omega_1^2 + i\tau_t \omega}\right), \qquad (2)$$

$$\frac{1}{E_{eff1}} = \frac{1}{E_0}\left(1 - \frac{F_p \omega_1^2}{\omega^2 - \omega_1^2 + i\tau_p \omega}\right). \qquad (3)$$

Experiments and theories have demonstrated no interactions among the surface plasmon polaritons[35-38]. Based on the phenomenon that acoustic artificial meta-molecules can simultaneously produce negative mass density and bulk modulus[33] and inspired by the fact that the visible white light is formed by seven-color lights, we propose a "flute-like" acoustic meta-molecule cluster model, we also design seven acoustic double meta-molecules with different geometric dimensions to construct the cluster. The length ratios of the units are 1, 5/7, 4/7, 3.5/7, 3/7, 2.5/7, and 2/7. The aperture ratios are 1 and 2.

Our previous studies have concluded that intrinsic resonant frequencies, and the double-negative frequency ranges of meta-molecules are determined by the tube length and diameter of the side hole. That is to say, the working frequency of the metamaterial can be modulated by changing the dimension of the meta-molecule. The locally resonant essence[33, 39-41] of the present metamaterial guarantees an insignificant influence among adjacent units; therefore, a broadband double-negative metamaterial can be realized by stacking meta-molecules with different working frequencies. As discussed using simulations and experiments, the number and structure size of the units are finally determined for this model. A meta-molecule cluster is engineered by arranging 14 meta-molecules, which are divided into seven units, as shown in Fig. 1b. The intervals of the adjacent units in x- and y-axes are 1 and 2 mm, respectively.

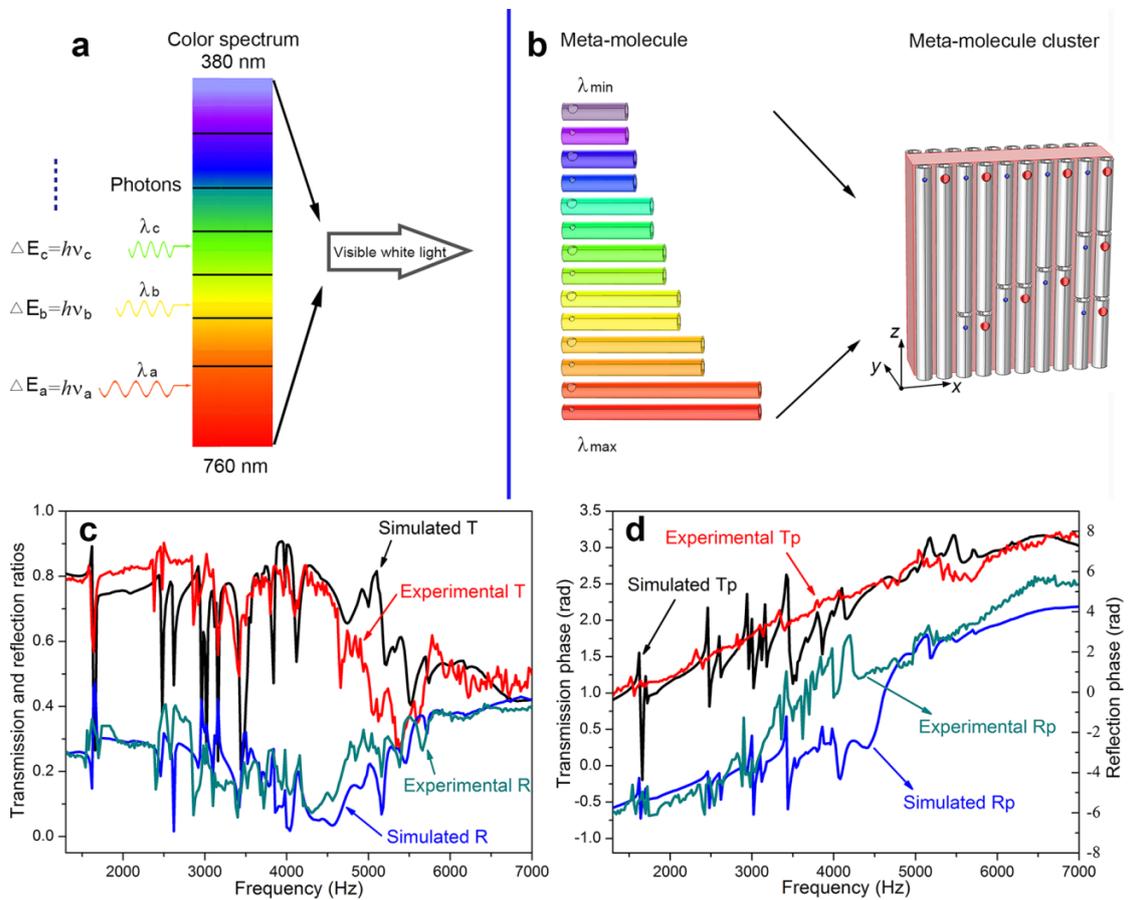

**Figure 1 Model and behavior of the acoustic meta-molecule cluster.** (a) Relationship between rainbow and visible white light. The frequency of a photon is determined by its energy, and lights with different frequencies have different colors. When we mix seven-color lights, visible white light appears. (b) Relationship between the meta-molecule units and the cluster. The units constructed with different dimensions generate acoustic surface

plasmon polaritons with different frequencies and anomalous acoustic properties. Seven meta-molecules are combined to form a cluster, with every meta-molecule containing two fine structures to realize broadband abnormal acoustic properties. Sound waves propagate along the positive direction of the y-axis. (c) Transmitted and reflected ratios of the metamaterial sample as a function of frequency. (d) Transmitted and reflected phases of sample. The black and red lines indicate the simulated and experimental transmittance curves, respectively. The blue and purple lines refer to the simulated and experimental reflectances, respectively.

The transmission and reflection behavior of the fabricated metamaterial sample is experimentally measured. The present cluster is also numerically simulated. The structural dimension of the simulated cluster is identical with that of the actual sample. A good agreement can be observed by comparing the experimental value and the simulated pattern, as shown in Fig. 1c, d, respectively. The transmittance curve shows a series of absorption peaks in a wide frequency range, within which phase shifts appear. This phenomenon is due to the fact that meta-molecule units with different dimensions correspond to different locally resonant frequencies. Moreover, no interaction occurs among adjacent units; each of them can resonate independently. The slight deviation of the transmission valleys within the experiments and simulations mainly stems from the machining error in the preparation of practical sample. The magnitude of the experimental transmission valley diverges from that of the simulated result because the environmental parameter set in the simulation does not match real circumstances well.

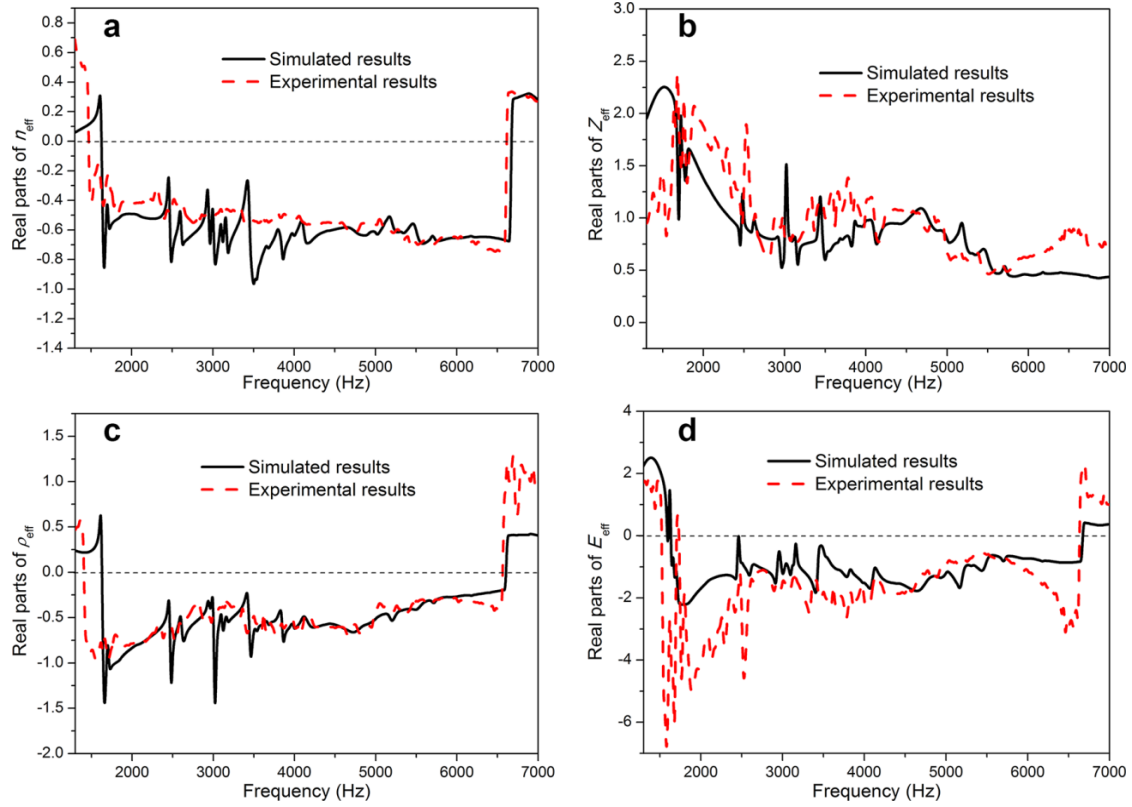

**Figure 2 Effective parameters of the metamaterial sample obtained by experiments and simulations.** (a) Effective refractive index. (b) Impedance. (c) Mass density. (d) Bulk modulus. The red and black lines represent the real parts of the results derived from the experimental and simulated data, respectively. The mass density and bulk modulus are negative in a very wide frequency band.

Figure 2 indicates the effective parameters (i.e., refractive index, impedance, mass density, and bulk modulus) as a function of frequency; they are derived from the transmission and reflection results based on the method in[42]. The real parts of the mass density and bulk modulus of the material are negative in a broad frequency range, which are 1.402 KHz to 6.56 KHz and 1.74 KHz to 6.635 KHz, respectively. The real part of refractive indices is negative from 1.472 KHz to 6.614 KHz.

A right triangle sample is fabricated based on the meta-molecule clusters. The refraction of the sample is measured experimentally in the frequency range of 0.8 KHz to 7.5 KHz (with 18 discrete frequencies). Figure 3a shows the schematic for the experimental measurement and the field distribution of sound pressure at 3.5 KHz. The transmitted and incident beams are on the same side of the normal. Based on the propagation direction of transmitted waves, the refraction of the sample is $n = -0.577$. Figure 3b exhibits the experimental and calculated results of the refractive indices of the metamaterial as a function of frequency. The measured refractive indices are negative within the frequencies between 1.238 and 6.214 KHz; this result matches the calculated results well.

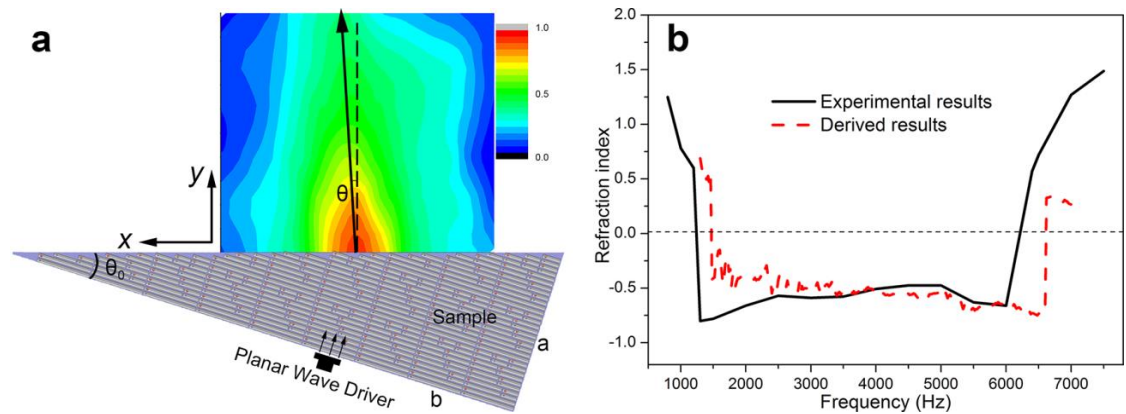

**Figure 3 Results of refraction for the metamaterial sample.** (a) Schematic for the measurement of refractive indics. $\theta_0 = 17.5°$, and the length of right-angle sides is a $\times$ b = 950 mm $\times$ 300 mm. The field pattern plotted in this figure is the experimental result at 3.5 KHz. (b) Relationship between refractive index and frequency. The black solid line indicates the measured results, and the red dashed line refers to the real part of the refractive indices derived from the transmission and reflection results.

One of the unique characteristics of double-negative metamaterials is the reversed Doppler effect. The Doppler effect of the proposed metamaterial is investigated experimentally. The experimental set-up is illustrated in Fig. 4a. The actual frequency of the moving source can be obtained by calculating the wave number in unit time, from which the Doppler shift of the constructed metamaterial can be derived. Experiments on the reversed Doppler effect are performed at frequencies ranging from 1.0 KHz to 6.7 KHz (16 discrete frequency points). Figure 4b displays the recorded signal by a stationary detector at 2.0 KHz. The moving source passes the sample centre at the middle of time axis (red dashed line); that is, the left and right sides of the red dashed line indicate the approaching and receding processes of the moving source, respectively. The measured Doppler shifts of the broadband double-negative sample versus frequencies are shown in Fig. 4c. Within a broad frequency band (1.5 KHz to 6.5 KHz), the Doppler shift of the metamaterial is not normal.

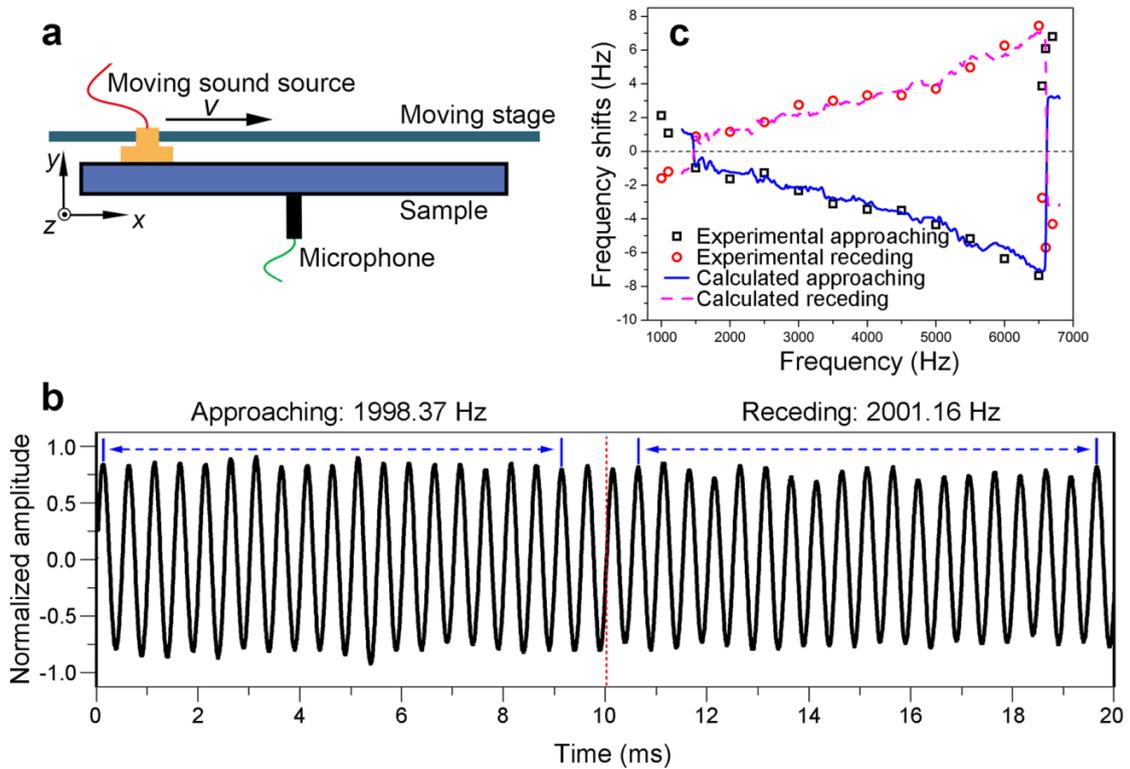

**Figure 4 Results of the reversed Doppler effect.** (a) Sketch map of the experimental setup for the Doppler shifts. (b) Oscillograms detected by a stationary microphone. The frequency of exciting sound wave is 2.0 KHz. When the source approaches the microphone, the detected frequency is reduced by 1.6 Hz; as the source recedes, the frequency increases by 1.2 Hz. (c) Doppler shifts of sample as a function of source frequency. The black boxes and red circles refer to the results of the approaching and receding processes, respectively. The positive value means

the detected frequency is larger than the source frequency, and the negative value implies the contrary situation. The blue solid line and the purple dashed line indicate the approaching and receding results derived from the refractive indices, respectively.

When sound waves propagate into the metamaterial from air, the relation between the sound velocity and refractive indices of the metamaterial is $n_1 v_1 = n_0 v_0$, where $n_1$ and $n_0$ are the refractive indices of the metamaterial and air, respectively; $v_1$ and $v_0$ represent the sound velocity of the metamaterial and air, respectively. As the observer is motionless and the source moves, the detected Doppler shift can be derived from the following equation[1]:

$$\Delta f = \left( \frac{v_1}{v_1 \mp v_s} - 1 \right) f_0 = \left( \frac{v_0/n_1}{v_0/n_1 \mp v_s} - 1 \right) f_0 = \left( \frac{v_0}{v_0 \mp v_s n_1} - 1 \right) f_0, \qquad (4)$$

where $\Delta f$ is the frequency shift; $v_s$ and $f_0$ are the speed and frequency of the sound source, respectively. Equation (4) indicates that if the speed of the sound source $v_s$ is fixed, then the Doppler shift is related only to the refractive index. The Doppler shifts calculated based on the derived refraction results are also displayed in Fig. 4c. The calculated results match the experimental results well.

Inspired by the fact that visible light is composed of multifrequency photons with seven different colors and based on acoustic surface plasmon polaritons and the previously presented meta-molecule fabricated by integrating a split hollow sphere meta-atom with negative bulk modulus and a hollow tube with negative mass density, we propose a model of "flute-like" acoustic double-meta-molecule cluster. We also provide a new approach to designing acoustic metamaterials with an arbitrarily broad band. The meta-molecule cluster consists of seven acoustic double meta-molecules with different dimensions. The length ratios of units are 1, 5/7, 4/7, 3.5/7, 3/7, 2.5/7, and 2/7. The aperture ratios are 1 and 2. Simulations and experiments verify that the metamaterial sample exhibits negative mass density in a frequency range of 1.402 KHz to 6.56 KHz, and a negative bulk modulus from 1.74 KHz to 6.635 KHz, realizing double negative in broad band. The measured refractive index of the right triangle sample is negative at frequencies ranging from 1.238 KHz to 6.214 KHz and changes continuously, which matches the derived results from the effective parameters and simulated results. The experiments demonstrate that the fabricated metamaterial sample exhibits a reversed Doppler effect from 1.5 KHz to 6.5 KHz, and the frequency shift increases with frequency. Assembling broadband double-negative metamaterials using double-meta-molecule clusters can be greatly tunable, which can be realized only by changing the structure size of the unit. This method of preparing an acoustic metamaterial paves a new way of designing and fabricating metamaterials with arbitrarily changing refractive index and broadband double-negative parameters. The method also exhibits great potential in such applications as broadband absorber and cloaking.

**Methods**

**Fabrication of Metamaterial samples.** The lengths of the tubes in a cluster are 98, 67, 55, 48, 41, 32.5, and 29 mm. The side hole is 5 mm away from one end of the tube, with diameters of 1 and 2 mm. The external and internal diameters of the tube are 7 and 5 mm, respectively. The material of tubes is plastic, which is hard enough for acoustic waves. The propagation medium of the acoustic wave was air. The clusters are pasted periodically on the front and back surfaces of a sponge substrate to form the metamaterial sample. The thickness of substrate is 20 mm. The sponge is a non-dispersive sound medium suitable for use as an acoustic substrate[22]. The dimension of the constructed metamaterial sample is 415 mm × 415 mm × 34 mm, which is used for the measurements of transmission, reflection, and Doppler shift. The dimension of the right-triangle sample is 950 mm × 300 mm × 34 mm, with $\theta_0$ of 17.5 °.

**Experimental facilities.** A plane wave driver (4510ND, BMS, Hannover, Germany) is connected to a signal generator (MC3242, BSWA, Beijing, China) and a power amplifier (PA50, BSWA, Beijing, China) to generate sine acoustic waves. A free field microphone (MPA416, BSWA, Beijing, China) is connected to a lock-in amplifier (SR830, SRS, Sunnyvale, USA) to record the amplitude and phase signals of sound wave. The samples are surrounded by sound absorbing materials to eliminate the scattered waves.

**Measurement methods of transmission and reflection of the sample.** The detailed measurement methods are presented in the reference 33.

**Measurement method of refraction of the right-triangle sample.** A planar wave driver is placed next to the sample and generates sound beams perpendicular to the interface. The incident beam forms an angle of 17.5 ° with the refraction surface. To map the sound field distribution of refracted waves on the *X–Y* plane, a microphone is fixed on a 3D translation stage with a scanning area of 200 mm $\times$ 50 mm.

**Measurement method of Doppler shifts of the metamaterial sample.** The sound source, launching sinusoidal acoustic signals, is mounted on a 1D motorized translation stage and moves along *X*-axis at a speed of 500 mm/s. A microphone is located at the center of the sample to record oscillograms as the source moves from one side of the sample to the other side (i.e., including the approaching and receding of the sound source from the observer). The loudspeaker and microphone are near the metamaterial surface, without contact.

**Acknowledgements**

This work was supported by the National Natural Science Foundation of China under Grant Nos. 11174234 and 51272215 and the National Key Scientific Program of China (under project No. 2012CB921503)

**Author contributions**

X.P.Z. and C.R.L. conceived the idea and designed the experiments; S.L.Z. and S.L. performed the major experiments; S.L.Z. performed the simulation study; S.L.Z. and X.P.Z. wrote the manuscript; S.L.Z. drafted the text and aggregated the figures; X.P.Z. and C.R.L. discussed the results and revised the manuscript.

**Additional information**

Supplementary information accompanies this paper at http://

**Competing financial interests**

The authors declare no competing financial interests.